\documentclass[12pt]{article}

\usepackage{epsfig}
\usepackage{scicite}
\usepackage{natbib}
\usepackage{subfigure,wrapfig}
\usepackage[]{sidecap}
\usepackage{caption}

\usepackage{times}

\newenvironment{sciabstract}{%
\begin{quote} \bf}
{\end{quote}}

\newcounter{lastnote}

\usepackage[T1]{fontenc}
\usepackage{mathptmx}
\usepackage{hyperref}

\hypersetup{
   colorlinks=true,
   citecolor=blue,
   linkcolor=blue,
   urlcolor=black
   }
   
\def\Rgp{R^{\star}}

\def\co2{CO$_2$}
\def\h2o{H$_2$O}
\def\ch4{CH$_4$}
\def\N2{N$_2$}

\def\nmix{N_{\mathrm{c}}}

\def\Fs{F_\star}

\def\grav{g}

\def\CLF{f_\mathrm{c}}

\def\temp{T}
\def\ps{p_\mathrm{s}}
\def\delps{\delta p_\mathrm{s}}
\def\press{p}

\def\delmad{\left. \frac{\partial \ln T}{\partial\ln p}\right |_{\mathrm{moist}}}

\def\mflux{W}

\def\mass{m}
\def\massg{m_\mathrm{g}}
\def\massgk{m_\mathrm{g}^{\kk}}

\def\delmassgk{\delta m_\mathrm{g}^{\kk}}

\def\qair{q_\mathrm{a}}
\def\rair{R_\mathrm{a}}
\def\cpair{c_\mathrm{p,a}}

\def\pvap{p_\mathrm{v}}

\def\qvap{q_\mathrm{v}}

\def\delqvap{\delta q_\mathrm{v}}

\def\cpvap{c_\mathrm{p,v}}

\def\delmvap{\delta m_\mathrm{v}}
\def\delmvapk{\delta \mass_\mathrm{v}^{\kk}}
\def\latent{L_\mathrm{v}}
\def\qsat{q_{\mathrm{s}}}
\def\psat{p_{\mathrm{s}}}

\def\qcon{q_\mathrm{c}}
\def\cpcon{c_\mathrm{p,c}}

\def\d{\mathrm{d}}
\def\i{\mathrm{i}}

\def\kk{\mathrm{k}}

\newcommand{\balign}[1]{
\begin{eqnarray}
#1
\end{eqnarray}}

\newcommand{\fracl}[2]{\raisebox{0.4ex}{$#1$} / \raisebox{-0.7ex}{$#2$}}

\newcommand{\eq}[1]{Eq.\,(\ref{#1})}

\newcommand{\sect}[1]{Sect.\,\ref{#1}}

\newcommand{\itref}[1]{$^{\ref{#1}}$}

\title{Increased insolation threshold for runaway greenhouse processes on Earth like planets}

\author{J\'er\'emy Leconte$^{1}$, Francois Forget$^{1}$, Benjamin Charnay$^{1}$,\\ Robin Wordsworth$^{2}$, and Aliz\'ee Pottier$^{1}$\\
\\
\normalsize{$^{1}$Laboratoire de M\'et\'eorologie Dynamique, Institut Pierre Simon Laplace, Paris, France,}\\
\normalsize{$^{2}$Department of Geological Sciences, University of Chicago, Chicago, IL, USA,}\\
\\
\normalsize{E-mail:  jeremy.leconte@lmd.jussieu.fr}
}

\date{}
\newpage

\include{journaux}

\begin{document} 


\baselineskip15pt


\maketitle 

\begin{sciabstract}

Because the solar luminosity increases over geological timescales, Earth climate is expected to warm, increasing water evaporation which, in turn, enhances the atmospheric greenhouse effect. Above a certain critical insolation, this destabilizing greenhouse feedback can "runaway" until all the oceans are evaporated\itref{Sim27}$^,$\itref{Kom67}$^,$\itref{Ing69}$^,$\itref{KPA84}. Through increases in stratospheric humidity, warming may also cause oceans to escape to space before the runaway greenhouse occurs\itref{Kas88}$^,$\itref{KRK13}. The critical insolation thresholds for these processes, however, remain uncertain because
they have so far been evaluated with unidimensional models that cannot account for the dynamical and cloud
feedback effects that are key stabilizing features of Earth's climate.
Here we use a 3D global climate model to show that the threshold for the runaway greenhouse is about 375 W/m$^2$, significantly higher than previously thought\itref{KRK13}$^,$\itref{GRZ13}. Our model is specifically developed to quantify the climate response of Earth-like planets to increased insolation in hot and extremely moist atmospheres. In contrast with previous studies, we find that clouds have a destabilizing feedback on the long term warming. However, subsident, unsaturated regions created by the Hadley circulation have a stabilizing effect that is strong enough to defer the runaway greenhouse limit to higher insolation than inferred from 1D models. Furthermore, because of wavelength-dependent radiative effects, the stratosphere remains cold and dry enough to hamper atmospheric water escape, even at large fluxes. 
This has strong implications for Venus early water history and extends the size of the habitable zone around other stars.

%because of these unsaturated regions, the stratosphere remains cold and dry enough to hamper atmospheric water escape, even at large fluxes. 
%This has strong implications for Venus early water history and extends the size of the habitable zone around other stars.

 \end{sciabstract}
 
Planetary atmospheres naturally settle into a thermal equilibrium state where their outgoing thermal emission balances the heating due to sunlight absorption. The resulting climate is stabilized by the fact that a temperature increase results in an enhanced thermal-emission cooling. When a condensable greenhouse gas is present at the surface, such as water on Earth, this stabilizing feedback is somewhat hampered by the destabilizing greenhouse feedback: evaporation, and thus water vapor greenhouse effect, increases with temperature, reducing the cooling. Fortunately, under present Earth conditions, this greenhouse feedback is both strong enough to maintain clement surface temperatures and weak enough for the climate to remain stable. 

When solar heating becomes stronger, however, water vapor can become abundant enough to make the atmosphere optically thick at all thermal wavelengths\itref{GRZ13}$^,$\itref{GW12}.
%, even in the so-called water vapor window around 10$\,\mu$m\itref{GW12}.
Thermal flux then originates from the upper troposphere only and reaches a maximum, $\sim282$\,W/m$^2$, independently from the surface temperature\itref{NHA92}. If the planet absorbs more than this critical flux, thermal equilibrium can be restored only by vaporizing all the water available and reaching high surface temperatures at which the surface starts to radiate at visible wavelengths\itref{KPA84}$^,$\itref{GRZ13}. This is the runaway greenhouse state. 

Because it has mostly been studied through unidimensional atmosphere models, the aforementioned mechanism strongly relies on the assumption that the troposphere is saturated in water vapor. Furthermore, by construction, these studies could not account for spatial inhomogeneities in insolation and in the resulting water vapor and cloud distributions\itref{KPA84}$^,$\itref{KRK13}$^,$\itref{GRZ13}. To overcome these limitations, we have developed a three-dimensional (3D) global climate model fit to describe hot atmospheres in which water vapor can become a dominant species\itref{WFS11}$^,$\itref{FWM13}$^,$\itref{LFC13}$^,$\itref{CFW13} (see Methods). The main challenges of such a model are threefold. First, the radiative transfer must be fast, yet able to describe accurately the spectroscopic properties of various gases in a wide temperature-pressure domain. Second, the modeling of the physical processes that are not specific to a given planet (convection, turbulence, etc) must rely on the fewest free parameters possible to ensure its validity in stringent conditions. Finally, for hot, moist atmospheres, the description of the water cycle (moist convection, cloud formation, etc) must take into account the fact that water can become a major constituent of the atmosphere. For these reasons, usual global climate models used to predict Earth climate are generally not suited for such studies.

\begin{figure}[htb] %  figure placement: here, top, bottom, or page
 \centering
  \subfigure{ \includegraphics[scale=.45,trim = 0.cm 0.cm .0cm 0cm, clip]{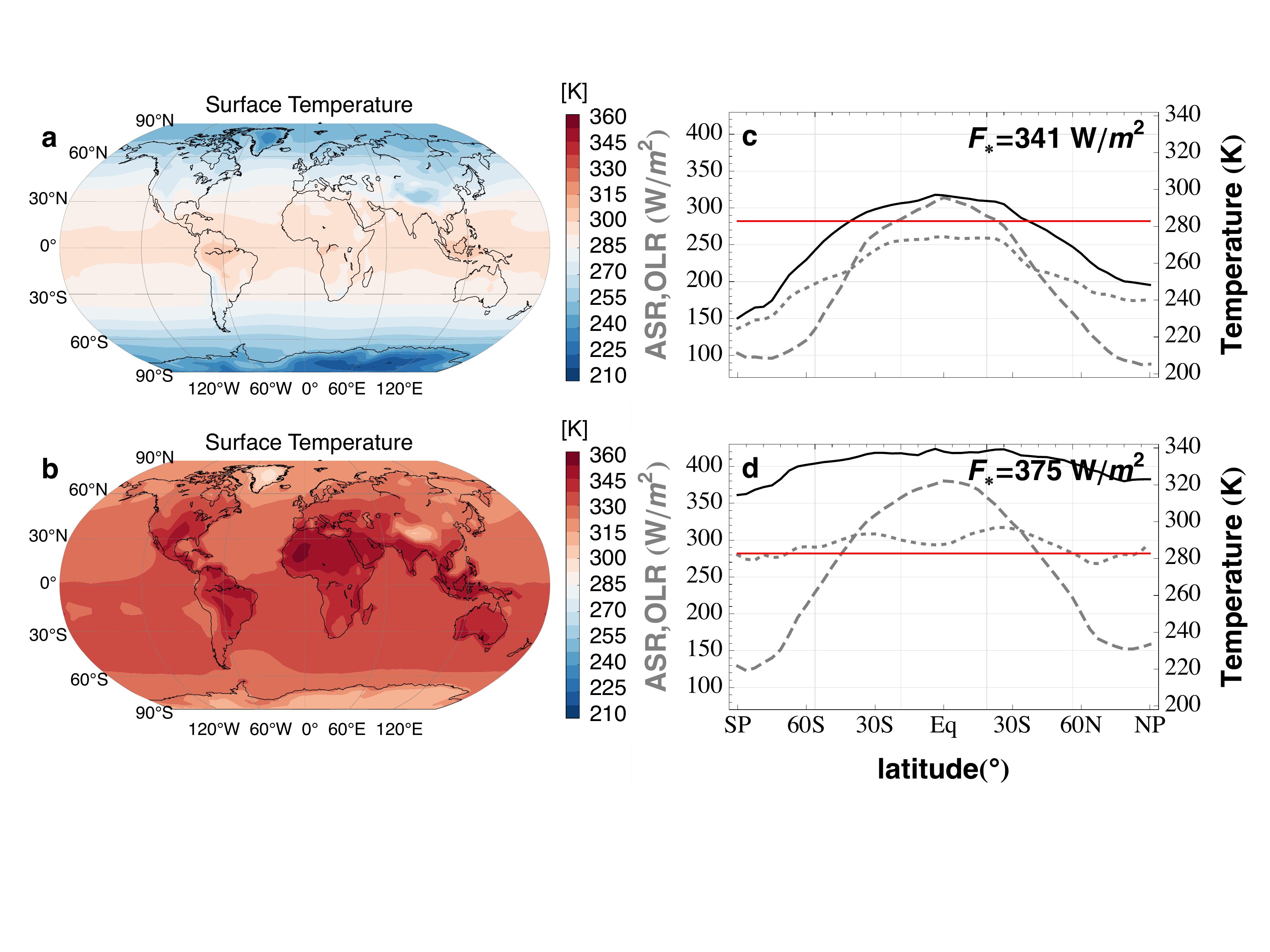} }
\vspace{-2cm}
\caption{
\textbf{Temperature and radiative budget for the Earth under two insolations.} Map of the annual mean surface temperature for the model corresponding to present Earth ($\Fs=341\,$W/m$^2$; a) and to a mean solar flux of 375\,W/m$^2$ (b), just before runaway greenhouse is triggered. c and d: Zonally and annually averaged surface temperature (solid black), and absorbed (gray dashed) and emitted (gray dotted) fluxes. The red horizontal line shows the radiation limit on the emitted flux for a saturated water atmosphere (282\,W/m$^2$). As visible in the hot case, unsaturated subtropical regions allow the atmosphere to emit more than this limit.
}
 \label{fig:Tmaps}
\end{figure}

Here we perform simulations of future Earth climate by running our baseline model for various (increasing) values of the solar constant until radiative balance is achieved.
%For each simulation, the model is run until long-term radiative balance is achieved.
For the current solar flux ($\Fs\approx341$\,W/m$^2$), our generic model reproduces the energetic budget and the characteristics of our climate\itref{CFW13} (see Fig.\,1). When the flux is increased, the planet undergoes a decrease in surface albedo which is due to the melting of the permanent polar ice caps and the reduced seasonal snow cover. Above $\sim$350\,W/m$^2$, only seasonal ice caps appear during the polar night. 
The amount of water vapor also increases. This results in a more efficient absorption of the incoming stellar light, but also in an enhanced greenhouse effect which tends to homogenize the surface temperatures. While continental surfaces can reach temperatures around 100$^\circ$C because of the intense solar and greenhouse heating, sea surface temperatures remain moderate with a small diurnal variation because they are thermodynamically controlled by latent heat cooling\itref{Pie10}. Finally, above $\sim$375\,W/m$^2$, no thermal equilibrium can be found. Although surface temperature increases with time, thermal emission reaches a limiting value. This is the onset of the runaway greenhouse instability.

This runaway greenhouse limit arises at higher fluxes than recently found by previous unidimensional studies\itref{KRK13}$^,$\itref{GRZ13} and confirmed by our 1D model (see Extended Data Fig.\,1). To understand the mechanisms deferring runaway greenhouse on actual planets, we first analyze the radiative effect of clouds. While 1D simulations cannot capture spatial variations in cloud distribution properly, it has been suggested that clouds should have a stabilizing feedback against runaway greenhouse effect\itref{KPA84}$^,$\itref{Kas88}$^,$\itref{KRK13}. This tentative conclusion was based on the fact that on present Earth, the net cloud radiative forcing is negative, meaning that the albedo increase due to low level clouds exceeds the greenhouse effect of high level clouds\itref{RCH89}. Several authors thus proposed that, because of the increased evaporation resulting from the warming, cloud thickness would increase and enhance cloud stabilizing effect\itref{KPA84}$^,$\itref{Kas88}$^,$\itref{KRK13}.

%\begin{figure}[htbp] %  figure placement: here, top, bottom, or page
% \centering
%  \subfigure{ \includegraphics[scale=.75,trim = 0.cm 0.cm .0cm 0cm, clip]{FigureFinalesNature/Figure2.pdf} }
%\vspace{-0.8cm}
%   \caption{
%\textbf{Evolution of the mean surface temperature, planetary albedo, and cloud radiative forcing with the mean solar incoming flux.} Curves start form present Earth conditions ($\sim 341\,$W/m$^2$). a) The solid curve stands for the average surface temperature in our 3D baseline model and the dashed curve for the temperature in the 1D cloud-free model. b) The solid line is the average surface albedo and the dashed line is the effective planetary albedo (or Bond albedo). c) Solid, dotted and dashed curves are the shortwave, long wave and net radiative cloud forcing in the baseline model.
%}
% \label{fig:meanquantities}
%\end{figure}

\begin{SCfigure}[1.][tbp]
  \centering
    \includegraphics[width=0.65\textwidth]{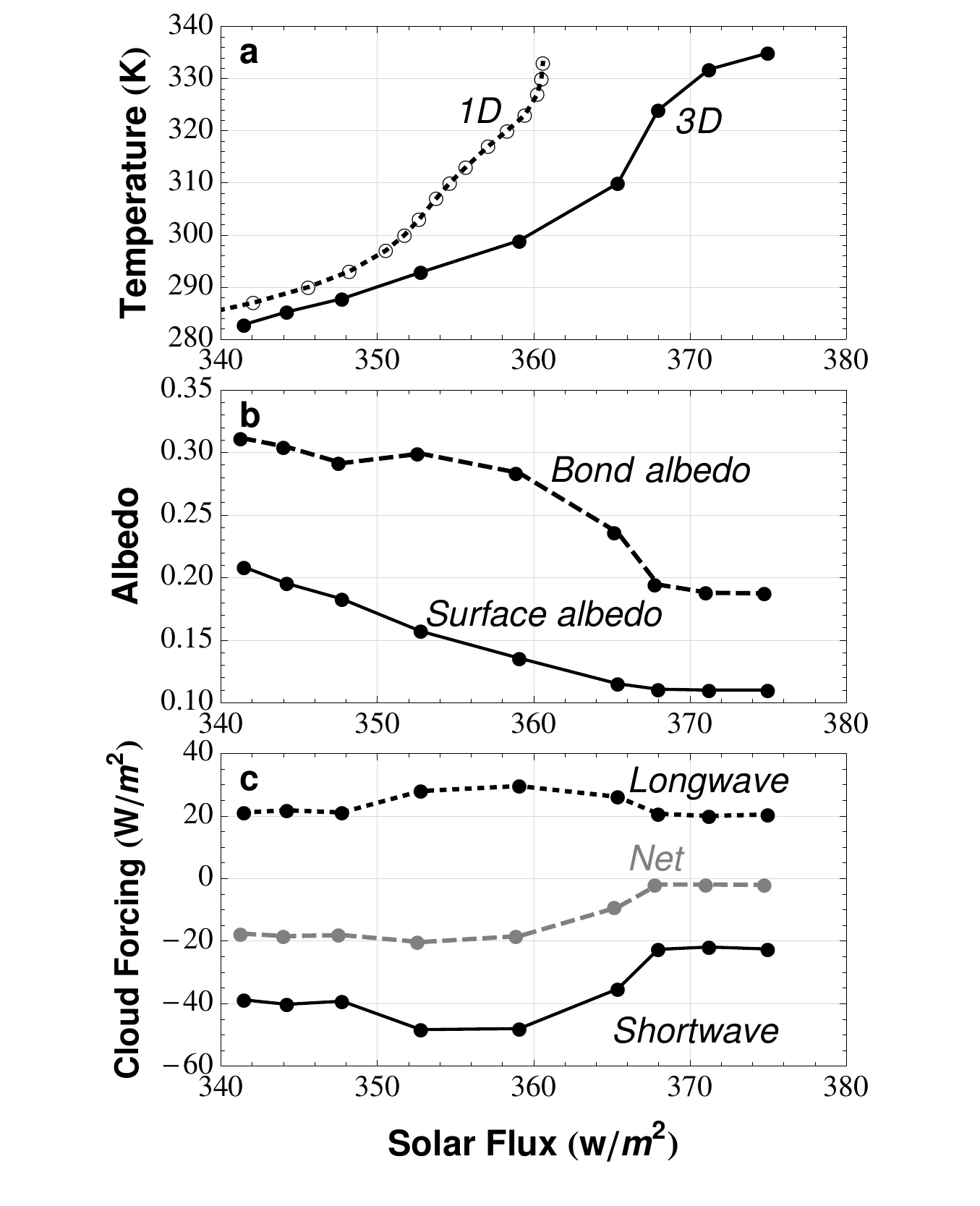}
    \caption{
    \textbf{Evolution of the mean surface temperature, planetary albedo, and cloud radiative forcing with the mean solar incoming flux.} Curves start form present Earth conditions ($\sim 341\,$W/m$^2$). a) The solid curve stands for the average surface temperature in our 3D baseline model and the dashed curve for the temperature in the 1D cloud-free model. b) The solid line is the average surface albedo and the dashed line is the effective planetary albedo (or Bond albedo). c) Solid, dotted and dashed curves are the shortwave, long wave and net radiative cloud forcing in the baseline model.
}
  \label{fig:1D3D}
\end{SCfigure}

As shown in Fig.\,2c where we show the evolution of the radiative cloud forcing with insolation, our simulations suggest the opposite.
This is due to a displacement of the cloud formation region toward higher altitudes (see Fig.\,3). As a result, the temperature at the mean cloud emission level increases much less with insolation than the surface temperature. Even though the cloud optical depth increases, the greenhouse feedback of the clouds exceeds their albedo effect. At higher fluxes, clouds become thinner, which reduces both longwave and shortwave radiative cloud forcing. For the reason described above, however, the greenhouse effect of clouds prevails and the net radiative forcing tends to cancel out.

There are several reasons for the cloud vertical displacement. First, moist convection, and Hadley circulation become more intense as the insolation increases, extending the troposphere.
% as seen in other 1D simulations\itref{KPA84}$^,$\itref{KRK13}.
Second, to form and persist, clouds need to be able to get rid of the latent heat released during condensation. Because of the infrared opacity increase of the atmosphere, the level at which clouds can efficiently cool radiatively rises. This may explain both the progressive disappearance of low level clouds and the small cloud-deck temperature change seen in Fig.\,3.

\begin{figure}[tbp] %  figure placement: here, top, bottom, or page
 \centering
  \subfigure{ \includegraphics[scale=.8,trim = 0.cm 0.cm .0cm 0cm, clip]{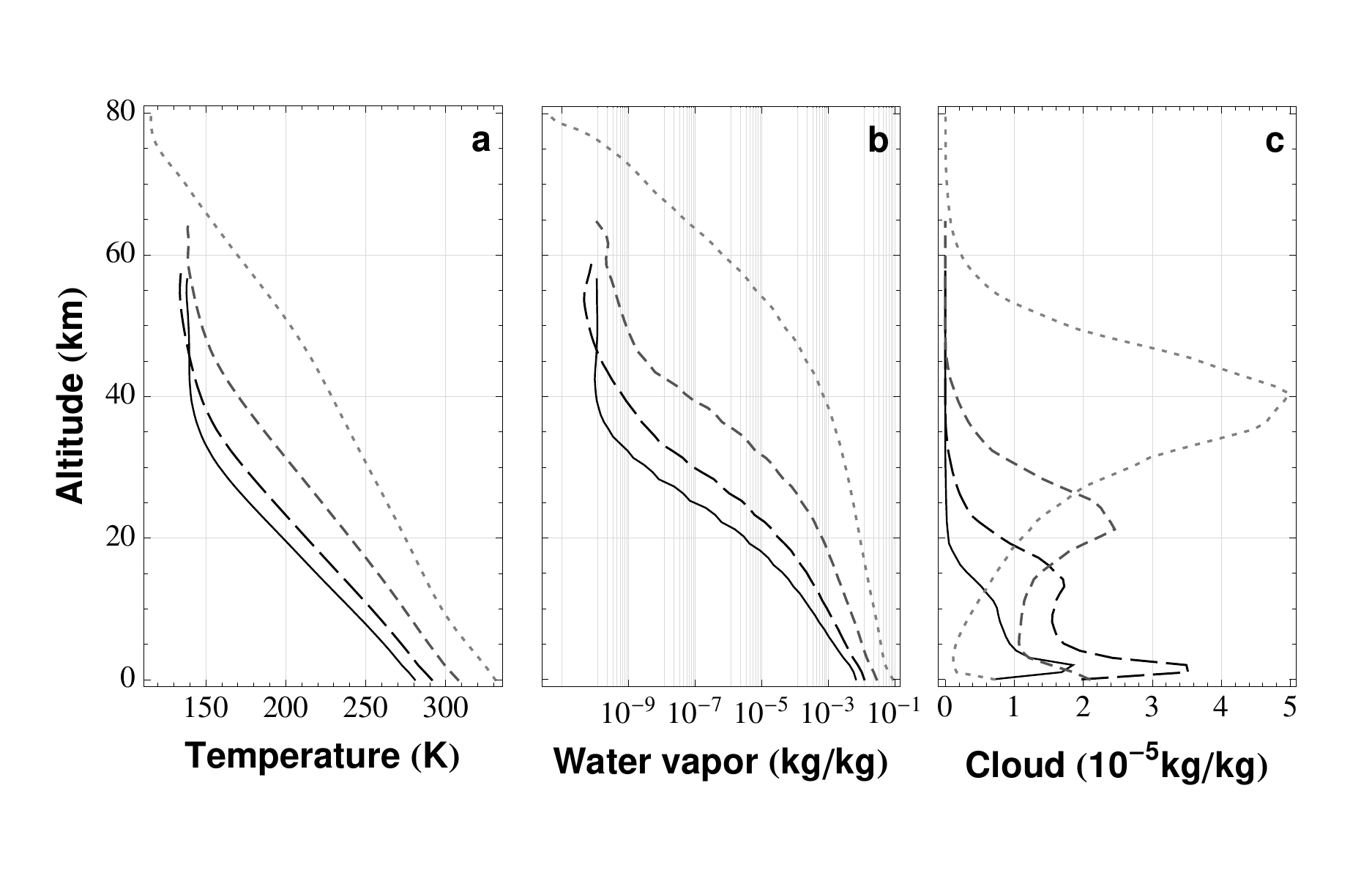} }
\vspace{-1.5cm}
 \caption{
\textbf{Evolution of globally averaged vertical profiles}. Mean vertical profiles of temperature (a), water vapor (b), and condensed water mixing ratios (c; in kg per kg of moist air) for 4 different insolations (solid: 341\,W/m$^2$; long dashed: 353\,W/m$^2$; dashed: 365\,W/m$^2$; dotted: 375\,W/m$^2$).
}
 \label{fig:profiles}
\end{figure}

Although the tendencies described above should be robust, the actual precise value of the cloud radiative forcing does depend on the assumptions made on the cloud microphysics. In our baseline model, for instance, we assume that the number density of cloud condensation nuclei, i.e. the number of cloud particles, per unit mass of air remains fixed to a value that is representative of modern Earth. As is expected, a larger mass mixing ratio of condensed water thus entails bigger cloud particles which precipitate more easily and have a smaller radiative effect. To explore the possible behavior of clouds and bracket the reality, we have conducted another set of simulations where it is the radii of cloud particles which is kept constant. This assumption results in smaller radii, and thus overestimates cloud optical depths, and shortwave and longwave cloud forcing. As can be seen in Extended Data Fig.\,2, longwave and shortwave forcing indeed increase continuously with insolation. However, for the reasons mentioned above, the greenhouse effect of clouds eventually overcomes the albedo effect. Therefore, the fact that the cloud feedback is destabilizing under an extreme insolation seems a robust conclusion which comes to support results obtained in the context of anthropogenic global warming\itref{SH06}.
%as long as cloud form over the whole planet as on Earth\itref{YCA13}. 
%Indeed, if the planet were in a synchronous or slow rotation, high clouds would preferentially form over the sunlit hemisphere, maximizing their sunshade effect while leaving the nightside free to radiate in the infrared\itref{YCA13}. 

If cloud do not help stabilize climate against a solar flux increase, atmospheric dynamics does. This is due to the fact that when a parcel of moist air is heated without any source of moisture, its water vapor pressure decreases relative to the saturation pressure. As a result, Earth troposphere itself is not saturated everywhere, unlike what is often assumed in 1D models. Sub tropical regions receiving hot air from the Hadley cell that has been dried during its ascent in the tropics and compressed adiabatically during its descent, are a perfect example\itref{SIT10} (see Fig.\,4). As they stabilize Earth tropics today, such dynamically unsaturated regions where water vapor greenhouse effect is reduced stabilize climate against runaway greenhouse by playing the role of "radiative fins" where the emission can exceed the maximum emission for a saturated atmosphere\itref{Pie95}$^,$\itref{ITN02} (see Fig.\,1). 

Furthermore, as seen in Fig.\,4, these unsaturated regions have the interesting property of extending upward and poleward with the Hadley cell (see Fig.\,4), and to get dryer when the insolation is increased\itref{SIT10} (see Extended Data Figs. 3 and 4). Such unsaturated regions, that cannot be predicted by a 1D model, ensure that the infrared photosphere remains always lower than in the saturated case, yielding a more efficient cooling to space. This stabilizing feedback is the main reason why the runaway greenhouse limit predicted by 3D simulations is closer to the Sun than previously found. And it would be even closer if it were not for the positive cloud feedback (see Methods).

\begin{figure}[htbp] %  figure placement: here, top, bottom, or page
 \centering
  \subfigure{ \includegraphics[scale=.8,trim = 0.cm 0.cm .0cm 0cm, clip]{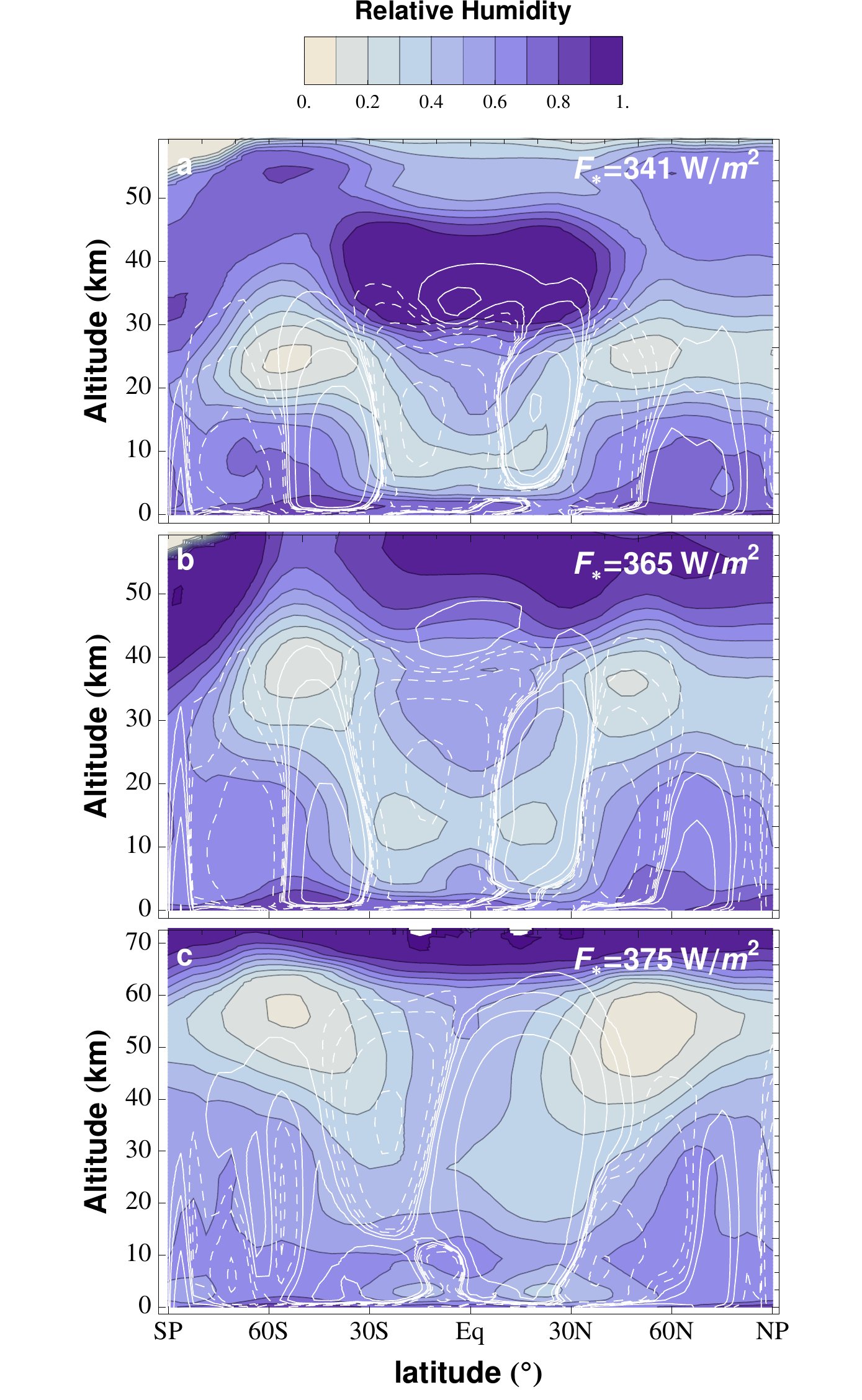} }
 \caption{
\textbf{Meridional distribution of relative humidity}. Zonally and annually averaged relative humidity (see color bar) for models receiving a stellar flux equal to 341, 365 and 375\,W/m$^{2}$ (from panel a to c). White contours are streamlines showing the Hadley circulation (solid and dashed curves depict clockwise and counter-clockwise rotation, respectively).
}
 \label{fig:RHmap}
\end{figure}

With this new estimation, the inner edge of the habitable zone for Earth like planet in the Solar System is pushed inward to $\sim$0.95\,AUs which means that the Earth should not enter a runaway greenhouse state before at least 1 billion years\itref{Gou81}. However, the question of whether or not the warming-induced increase of the stratospheric temperature and humidity could cause the loss of the oceans by atmospheric escape before the runaway greenhouse is triggered, as suggested by some 1D studies\itref{KPA84}$^,$\itref{KRK13}, is still subject to debate. Our simulations tend to invalidate this scenario. This is due to both non-gray radiative effects\itref{Pie10}$^,$\itref{AG72}$^,$\itref{WP13} (see Methods) and unsaturated regions that flatten the thermal profile in the troposphere.
% Heat transport toward higher latitudes also tend to cool the stratosphere in tropical regions.
As a result, stratospheric temperature are much colder than anticipated and stratospheric humidity cannot reach the threshold needed for efficient water photodissociation and hydrogen escape to space (see Fig.\,3). 

While extending the habitable zone\itref{KWR93} towards the star, more importantly, our results highlight the fact that global climate models are needed to understand subtle climate feedbacks resulting from the inhomogeneous insolation of planetary surfaces that cannot be modeled in 1D. Although adding complexity and uncertainties, these subtle effects must be accounted for in modeling real planets, and especially tidally-synchronized exoplanets where they are even more pronounced\itref{LFC13}$^,$\itref{Pie11}$^,$\itref{YCA13}. In particular, while our simulation suggest that Venus, if it had Earth rotation rate, would have been in a runaway greenhouse state since its formation, its slow retrograde motion (because it results in a long Solar day), or a small water inventory\itref{LFC13}$^,$\itref{AAS11}, could completely change the picture.

%At first (insolation below 360\,W/m$^2$), even if the albedo effects of clouds in the optical part of the spectrum increases, it is balanced by an increase in the greenhouse effect in the infrared, and the absolute value of the net radiative cloud forcing does not change or slightly decreases with increasing flux.

% This is due to the fact that when a parcel of moist air is heated without any source of moisture (during an adiabatic compression along the subsiding branch of the Hadley cell for instance), its water vapor pressure decreases relative to the saturation pressure   On present Earth, the inter tropical convergence zone could not reach radiative equilibrium if the Hadley circulation was/were not carrying  

%To understand the mechanisms deferring runaway greenhouse effect in 3D simulations, we have run our model in various idealized 3D configurations and a 1D mode similar to previous studies to disentangle differences in the spectroscopic data used from dynamical and cloud effects (see Supplementary Informations). As shown in \fig{?}, at all fluxes, the 1D model always predicts higher surface temperatures. 

%\bigskip
%\clearpage

\section*{Methods Summary}

Our simulations were performed with an upgraded version of the LMD generic global climate model (GCM) specifically developed for the study of extrasolar planets\itref{WFS11}$^,$\itref{LFC13}$^,$\itref{SWF11} and paleoclimates\itref{FWM13}$^,$\itref{WFM13}. The model uses the 3D dynamical core of the LMDZ Earth GCM used in IPCC studies\itref{HMB06}, based on a finite-difference formulation of the primitive equations of geophysical fluid dynamics. 

General physical processes relevant for present Earth - including ground thermal inertia and albedo (see Extended Data Fig.\,5), turbulent transport, dry convection, evaporation/condensation and precipitations - are parametrized in the most physically based way to ensure the robustness of the model under extreme conditions\itref{CFW13}. For the present study, special care has been taken to treat properly situation where water vapor can become a major constituent of the atmosphere. As detailed in the online-only Methods, the radiative transfer and moist convection scheme have been validated against recent 1D results in the hot, pure water atmosphere regime (see Extended Data Fig.\,1). A numerical scheme accounting for change in atmospheric mass and surface pressure with water vapor evaporation/condensation has also been implemented.

\clearpage

\section*{ References and Notes}

\begin{enumerate}
\item \label{Sim27} Simpson, G. C. Some studies in terrestrial radiation. \textit{Mem. Roy. Met. Soc.} (1927).
\item \label{Kom67} Komabayashi, M. Discrete equilibrium temperatures of a hypothetical planet with the atmosphere and the hydrosphere of one component-two phase system under constant solar radiation. \textit{J. Meteor. Soc. Japan}, \textbf{45}, 137-138 (1967).
\item \label{Ing69} Ingersoll, A. P. The Runaway Greenhouse: A History of Water on Venus. \textit{Journal of Atmospheric Sciences}, \textbf{26}, 1191-1198 (1969).
\item \label{KPA84} Kasting, J. F., Pollack, J. B. and Ackerman, T. P. Response of earth's atmosphere to increases in solar flux and implications for loss of water from Venus. \textit{Icarus}, \textbf{57}, 335-355 (1984).
\item \label{Kas88} Kasting, J. F. Runaway and moist greenhouse atmospheres and the evolution of earth and Venus. \textit{Icarus}, \textbf{74}, 472-494 (1988).
\item \label{KRK13} Kopparapu, R. K., Ramirez, R., Kasting, J. F., Eymet, V., Robinson, T. D., Mahadevan, S., Terrien, R. C., Domagal-Goldman, S., Meadows, V. and Deshpande, R. Habitable Zones around Main-sequence Stars: New Estimates. \textit{Astrophys. J.}, \textbf{765}, 131 (2013).
\item \label{GRZ13} Goldblatt, C., Robinson, T. D., Zahnle, K. J., and Crisp, D. Low simulated radiation limit for runaway greenhouse climates. \textit{Nature Geoscience}, \textbf{6}, 661-667 (2013).
\item \label{GW12} Goldblatt, C. \& Watson, A. J. The runaway greenhouse: implications for future climate change, geoengineering and planetary atmospheres. \textit{Royal Society of London Philosophical Transactions Series A}, \textbf{370}, 4197-4216 (2012).
\item \label{NHA92} Nakajima, S., Hayashi, Y.-Y. and Abe, Y. A study on the ?runaway greenhouse effect? with a one-dimensional radiative-convective equilibrium model. \textit{Journal of Atmospheric Sciences}, \textbf{49}, 2256-2266 (1992).
\item \label{WFS11} Wordsworth, R. D., Forget, F., Selsis, F., Millour, E., Charnay, B. and Madeleine, J.-B. Gliese 581d is the First Discovered Terrestrial-mass Exoplanet in the Habitable Zone. \textit{Astrophys. J.}, \textbf{733}, L48 (2011).
\item \label{FWM13} Forget, F., Wordsworth, R., Millour, E., Madeleine, J.-B., Kerber, L., Leconte, J., Marcq, E. and Haberle, R. M. 3D modelling of the early martian climate under a denser CO2 atmosphere: Temperatures and CO2 ice clouds. \textit{Icarus}, \textbf{222}, 81-99 (2013).
\item \label{LFC13} Leconte, J., Forget, F., Charnay, B., Wordsworth, R., Selsis, F., Millour, E., and Spiga, A. 3D climate modeling of close-in land planets: Circulation patterns, climate moist bistability, and habitability. \textit{Astron. Astrophys.}, \textbf{554}, A69 (2013).
\item \label{CFW13} Charnay, B., Forget, F., Wordsworth, R., Leconte, J., Millour, E., Codron, F., Spiga, A. Exploring the faint young Sun problem and the possible climates of the Archean Earth with a 3D GCM. \textit{J. Geophys. Res. (Atmospheres)}, \textbf{118}, 414-431 (2013).
\item \label{Pie10} Pierrehumbert, R. T. \textit{Principles of Planetary Climate} (2010).
\item \label{RCH89} Ramanathan, V., Cess, R. D., Harrison, E. F., Minnis, P., Barkstrom, B. R., Ahmad, E., Hartmann,  D. Cloud-Radiative Forcing and Climate: Results from the Earth Radiation Budget Experiment. \textit{Science}. \textbf{243}, 57-63 (1989).
\item \label{SH06} Soden, B. J. \& Held, I. M. An Assessment of Climate Feedbacks in Coupled Ocean Atmosphere Models. \textit{Journal of Climate}, \textbf{19}, 3354, 2006.
\item \label{SIT10} Sherwood, S. C., Ingram, W., Tsushima, Y., Satoh, M., Roberts, M., Vidale, P. L., and O'Gorman, P. A. Relative humidity changes in a warmer climate. \textit{J. Geophys. Res. (Atmospheres)}, \textbf{115}, 9104 (2010).
\item \label{Pie95} Pierrehumbert, R. T. Thermostats, Radiator Fins, and the Local Runaway Greenhouse. \textit{Journal of Atmospheric Sciences}, \textbf{52}, 1784-1806 (1995).
\item \label{ITN02} Ishiwatari, M., Takehiro, S.-I., Nakajima, K., Hayashi, Y.-Y. A Numerical Study on Appearance of the Runaway Greenhouse State of a Three-Dimensional Gray Atmosphere. \textit{Journal of Atmospheric Sciences}, \textbf{59}, 3223-3238 (2002).
\item \label{Gou81} Gough, D. O. Solar interior structure and luminosity variations. \textit{Sol. Phys.}, \textbf{74}, 21-34 (1981).
\item \label{AG72} Arking, A., Grossman, K. The influence of line shape and band structure on temperatures in planetary atmospheres. \textit{Journal of Atmospheric Science}, \textbf{29}, 937-949, (1972).
\item \label{WP13} Wordsworth, R. \& Pierrehumbert, R. T. Water loss from terrestrial planets with CO2-rich atmospheres. ArXiv e-prints (2013).
\item \label{KWR93} Kasting, J. F., Whitmire, D. P., and Reynolds, R. T. Habitable Zones around Main Sequence Stars. \textit{Icarus}, \textbf{101}, 108-128 (1993).
\item \label{Pie11} Pierrehumbert, R. T. A Palette of Climates for Gliese 581g. \textit{Astrophys. J.}, \textbf{726}, L8 (2011).
\item \label{YCA13} Yang, J., Cowan, N. B., and Abbot, D. S. Stabilizing Cloud Feedback Dramatically Expands the Habitable Zone of Tidally Locked Planets. \textit{Astrophys. J.}, \textbf{771}, L45 (2013).
\item\label{AAS11} Abe, Y., Abe-Ouchi, A., Sleep, N. H., and Zahnle, K. J. Habitable Zone Limits for Dry Planets. \textit{Astrobiology}, \textbf{11}, 443-460 (2011).
\item \label{SWF11} Selsis, F., Wordsworth, R. D., and Forget, F. Thermal phase curves of nontransiting terrestrial exoplanets. I. Characterizing atmospheres. \textit{Astron. Astrophys.}, \textbf{532}, A1 (2011).
\item \label{WFM13} Wordsworth, R., Forget, F., Millour, E., Head, J. W., Madeleine, J.-B., and Charnay, B. Global modelling of the early martian climate under a denser CO2 atmosphere: Water cycle and ice evolution. \textit{Icarus}, \textbf{222}, 1-19 (2013).
\end{enumerate}

\subsubsection*{Acknowledgments}
We thank our referees, J. Kasting and Y. Abe, for their thorough review, and A. Spiga and F. Selsis for discussions. This work was supported by grants from Région Ile-de-France
%J.L. received support from the Domaine d'Intérêt Majeur d'Île de France: Astrophysique et Condition d'Apparition de la Vie (DIM ACAV)

\subsubsection*{Author contributions}
J.L. developed the "high temperature/humidity" version of the generic global climate model, performed the calculations, and led the analysis and writing of the results. F.F. initiated the development of the generic global climate model and provided critical advice during analysis and writing. B.C worked on the development of the model and helped performing the comparison with current Earth climatology. RW developed the original version of the generic model, including the correlated-k radiative transfer scheme. A.P. performed comparison runs and sensitivity studies. All the authors commented on the manuscript. 

\subsubsection*{Author Information}
Author Information Reprints and permissions information is available at www.nature.com/reprints. The authors declare no competing financial interests. Readers are welcome to comment on the online version of the paper. Correspondence and requests for materials should be addressed to J.L. (jeremy.leconte@lmd.jussieu.fr).

\clearpage

\clearpage

\section*{Methods}

\section{Numerical climate model}

Our simulations were performed with an upgraded version of the LMD generic global climate model (GCM) specifically developed for the study of extrasolar planets\itref{WFS11}$^,$\itref{LFC13}$^,$\itref{SWF11} and paleoclimates\itref{FWM13}$^,$\itref{WFM13}. The model uses the 3D dynamical core of the LMDZ Earth GCM used in IPCC studies\itref{HMB06}, based on a finite-difference formulation of the primitive equations of geophysical fluid dynamics. 

\subsection{Generalities}

A horizontal resolution of $64\times48$, corresponding to resolutions of 3.75$^\circ$ latitude by 5.625$^\circ$ longitude, is used for the simulations. The vertical grid uses hybrid coordinates: a terrain-following $\sigma$ coordinate system in the lower atmosphere ($\sigma$ being equal to the pressure divided by the surface pressure), and pressure levels in the upper atmosphere. In this work, we used 30 layers, with the lowest mid-layer level at 4 m and the top mid layer level at 1\,mb. The dynamical core is called every 90\,s, the tendencies given by the physical parametrizations are updated every 15\,min and the radiative transfer is called every 90\,min.

The boundary layer dynamics are accounted for by the unstationary 2.5-level closure scheme of ref \ref{MY82} , plus a convective adjustment which rapidly mixes the atmosphere in the case of dry unstable temperature profiles. Turbulence and convection mix energy (potential temperature), momentum (wind), and water (condensed and gaseous). A standard roughness coefficient of $z_0$ = 10$^{-2}$ m is used for both rocky and ocean surfaces for simplicity.

The evolution of surface temperature is governed by the balance between radiative, latent, and sensible heat fluxes (direct solar insolation, thermal radiation from the atmosphere and the surface, and turbulent and latent heat fluxes; see \sect{sec:water_cycle}) and thermal conduction in the soil. The parameterization of this last process is based on an 13-layer soil model solving the heat diffusion equation using finite differences. The depth of the layers were chosen to capture diurnal thermal waves as well as the deeper annual thermal wave. A vertically homogeneous soil is assumed. The thermal inertia of continental surface is set to 2\,000\,Js$^{-1/2}$m$^{-2}$K$^{-1}$. To model the high thermal inertia due to mixing in the upper layers of the oceans, the thermal inertia of the oceans is set to 18\,000\,Js$^{-1/2}$m$^{-2}$K$^{-1}$. The albedo surface map used in this work is shown in Extended Data Fig.\,5

\subsection{Radiative transfer}

The method used to produce our radiative transfer model is similar to\itref{WFS11}$^,$\itref{LFC13}. For a gaseous composition similar to the Earth (1\,bar of \N2 with 376\,ppmv of \co2 and a variable amount of water vapor; CH$_4$, O$_2$, and O$_3$ have been discarded for more generality), we computed high-resolution spectra over a range of temperatures and pressures using the HITRAN 2008 database\itref{RGB09}. For this study we used temperature and pressure grids with values $T = \{110,170,...,710\}\, \mathrm{K}, p = \{10^{-3},10^{-2},...,10^{5}\}\, \mathrm{mbar}$. The \h2o volume mixing ratio could vary in the range $\{10^{-7},10^{-6},...,1\}$. The \h2o lines were truncated at 25\,cm$^{-1}$, while the water vapor continuum was included using the CKD model\itref{CKD89}. We also account for opacity due to \N2-\N2 collision-induced absorption\itref{BF86}$^{,}$\itref{RGR11}.

The correlated-k method was then used to produce a smaller database of coefficients suitable for fast calculation in a GCM. Thanks to the linearity of the Schwarzschild equation of radiative transfer, the contribution of the thermal emission and downwelling stellar radiation can be treated separately, even in the same spectral channel. We therefore do not assume any spectral separation between the stellar and planetary emission wavelengths. For thermal emission, the model uses 19 spectral bands, and the two-stream equations are solved using the hemispheric mean approximation\itref{TMA89}. Absorption and scattering of the downwelling stellar radiation is treated with the $\delta$-Eddington approximation within 18 bands. Sixteen points are used for the $\bar{g}$-space integration, where $\bar{g}$ is the cumulated distribution function of the absorption data for each band.
Rayleigh scattering by N$_2$ and H$_2$O molecules is included using the method described in ref \ref{HT74} with updated cross section for water\itref{Buc95}. %Both infrared and optical scattering by clouds is taken into account using\itref{???}.

\subsection{Water cycle}\label{sec:water_cycle}

In the atmosphere, we follow the evolution of water in its vapor and condensed phases. These tracers are advected by the dynamical core, mixed by turbulence and dry and moist convection. Much care has been devoted to develop a robust and numerically stable water cycle scheme that is accurate both in the trace gas (water vapor mass mixing ratio $\qvap\ll 1$) and dominant gas ($\qvap\sim 1$) limit.
In particular, the atmospheric mass and surface pressure variation (and the related vertical transport of tracers through pressure levels) due to any evaporation/sublimation/condensation of water vapor is taken into account (see next section).

\subsubsection{Cloud formation}

Cloud formation is treated using the prognostic equations of ref \ref{LL91}. For each column and level, this scheme provides the cloud fraction $\CLF$ and the mass mixing ratio of condensed water $\qcon$, which are both functions of $\qvap$ and the saturation vapor mass mixing ratio $\qsat$. In addition, when part of a column reaches both 100\% saturation and a super-moist-adiabatic lapse rate, moist convective adjustment is performed following ref \ref{MW67}, and the cloud fraction is set to unity. This moist convection scheme has been chosen instead of more refined ones because it is more robust for a wide range of pressure, at the cost of giving enhanced precipitations at the equator\itref{Fri07}. Furthermore, the moist-adiabat lapse rate has been generalized to account for the fact that water can be a dominant species. This yields 
\balign{\label{lapse_rate}
\delmad=\frac{\press}{\press-\pvap}\left(\frac{\left(1-\qvap \right)\rair+ \qvap\frac{\latent}{T}}{ \qvap \cpvap+\qair \cpair +\qcon\cpcon + \qvap \frac{\latent}{T} \frac{\press}{\press-\pvap}\frac{\d \ln \psat}{\d \ln T} }\right),}
where $p_\i$, $M_\i$, $c_\mathrm{p,i}$, and $R_\i\equiv \Rgp/M_\i$ are respectively the pressures, molar masses, specific heat capacity at constant pressure, and specific gas constants of the various phases (non condensable gas, or air, denoted with a subscript a; condensible gas, or vapor, denoted by v; condensed material, denoted by c); $\press$ and $\temp$ being the total pressure and temperature, $\Rgp$ the molar ideal gas constant, $\latent$ the specific latent heat of vaporization and $\psat$ the water vapor saturation pressure.

In our baseline model, the liquid/icy water (depending on the temperature) is assumed to condense on a number $\nmix$ of activated cloud condensation nuclei (CCN) per unit mass of moist air. This number density of CCNs is assumed to be spatially uniform but to have a different value for liquid or ice cloud particles (the condensed water phase is determined by the local temperature). Indeed, on Earth, nucleation is found to be much less efficient in cold, high level clouds, resulting in larger particle radii in these ice clouds compared to liquid water clouds. This dichotomy is essential to recover the observed balance between longwave and shortwave cloud radiative forcings. To recover current Earth climate, the values used are $4\times10^6$\,kg$^{-1}$ for liquid water clouds and $2\times10^4$\,kg$^{-1}$ for ice cloud.

 Then, the effective radius of the cloud particles is given by
$r_\mathrm{eff}=\left(\fracl{3\,\qcon}{4\,\pi\,\rho_\mathrm{c}\,\nmix}\right)^{1/3}$, where $\rho_\mathrm{c}$ is the density of condensed water (10$^3$\,kg/m$^3$ for liquid and 920\,kg/m$^3$ for ice). In the special case where radii have been kept fixed, the values used where 12 and 38\,$\mu$m for water droplets and ice particles respectively. Precipitations are computed with the scheme given in ref \ref{BLB95}. Since this scheme explicitly considers the dependence on gravity, cloud particle radii, and background air density, it should remain valid for a wide range of situations (see ref \ref{CFW13} for details). Finally, the total cloud fraction of the column is assumed to be the cloud fraction of the level with the thickest cloud, and radiative transfer is computed both in the cloudy and clear sky regions. Fluxes are then linearly weighted between the two regions.

\subsubsection{Hydrology and evaporation}

The ground is modeled using a simple bucket model with a maximal water capacity of 150 kg/m$^2$. When the water amount exceeds this limit, the surplus is regarded as runoff and added to the oceans.
The effect of latent heat release during solidification/melting of snow/ice at the surface is taken into account.
On the surface, ice can also have a radiative effect by linearly increasing the albedo of the ground to $A_\mathrm{ice}=0.5$ until the ice surface density exceeds a certain threshold (here 30\,kg\,m$^{-2}$). 

Evaporation $E$ (in kg/m$^2$/s) is computed within the boundary layer scheme, using a bulk aerodynamic formula multiplied by a dryness coefficient ($\beta$ which is zero for a dry surface and linearly increases towards 1 when the water amount reaches 75\,kg/m$^2$ at the surface; $\beta$=1 over oceans). This yields
 \begin{equation}
E= \rho\, C \,V \beta \left[\qsat(T_\mathrm{surf})-\qvap^{1}\right]
 \end{equation}
where $ \rho$ is the mass density of air, $V$ the wind speed above the surface, $\qsat(T_\mathrm{surf})$ the water vapor mass mixing ratio at saturation at the surface, and $\qvap^{1}$ the mixing ratio in the first layer. The aerodynamic coefficient is given by $C=\kappa/\ln(1+z_1/z_0)$, where $\kappa=0.4$ is the Von Karman constant, $z_0$ the roughness and $z_1$ the altitude of the first level.

\subsection{A numerical scheme for atmospheric mass redistribution due to the condensation of a non-trace gas.}

In most GCM's, when a trace gas condenses in a grid box, the variation of the total gas mass is neglected and the mass mixing ratio variation is given by $\delqvap=\delmvap/\massg$. Because water vapor is not a trace gas in our simulations, we must take into account this effect as well as the change of surface pressure (total mass of the atmosphere) entailed by the massive evaporation/condensation that can take place both at the surface and aloft.
To that purpose, we developed a numerical scheme similar to the one used for \co2 clouds on Mars\itref{FML06}. This scheme can be used for any kind of condensing species without any assumptions on the mixing ratio of the vapor phase. The scheme is thus valid both in the trace and dominant gas limit as well as in intermediate regimes.

This is done in two steps. First, the various routine describing the water cycle (evaporation at the surface, moist convection, cloud condensation, precipitations) compute the mass of vapor that has been added at a level $\kk$, $\delmvapk$ (>0 when vapor is created\footnote{$\delmvap^0$ is the mass of water vapor evaporated directly at the ground}), assuming that the gas mass in the layer is constant (usual approximation). Second, the change in the total pressure and mass mixing ratios due to the change in the total mass of gas ($\delmassgk$) is computed as follows.

For an atmospheric column of N layers and area $A$, the change in the surface pressure is given by
\balign{\label{totpressvar}
\delps= \frac{\grav}{A}\sum_{\kk=0}^{\mathrm{N}} \delmvapk,
}
where $\grav$ is the gravity. For the computation in each cell, the difficulty comes from the fact that we do not use lagrangian coordinates. As a result, evaporation/condensation occurring both at the surface and aloft induces artificial mass fluxes through horizontal levels that must be accounted for.

For hybrid coordinates, as used in many GCMs, layer $\kk$ encompasses the matter between $p^{\kk-\frac{1}{2}}$ and $p^{\kk+\frac{1}{2}}$, where $p^{\kk}\equiv \sigma^{\kk} \ps + \gamma^{\kk}$. The so-called $\sigma$ coordinates are retrieved by setting $\sigma^{\kk}\equiv p^{\kk}/\ps$ and $\gamma^{\kk}=0$. Because both $\sigma^{\kk}$ and $\gamma^{\kk}$ are time independent, the variation in the gas mass in the layer during one timestep is given by 
\balign{\label{delmasssigma}
\delmassgk= \frac{A}{\grav}\left(  \delta p^{\kk-\frac{1}{2}} - \delta p^{\kk+\frac{1}{2}} \right)= \frac{A}{\grav}\left(  \sigma^{\kk-\frac{1}{2}} - \sigma^{\kk+\frac{1}{2}} \right) \delps.
}
Besides, the variation in the gas mass in layer $\kk$ is also linked to the mass flux through its interfaces ($\mflux^{\kk-\frac{1}{2}}$ being the mass flux in kg between level $\kk-1$ and $\kk$ counted positive when going upward) by
\balign{\label{delmassW}
\delmassgk= \delmvapk+\mflux^{\kk-\frac{1}{2}}-\mflux^{\kk+\frac{1}{2}}
}

Using Eqs\,(\ref{totpressvar}), (\ref{delmasssigma}) and (\ref{delmassW}), we can get a recursive formula for the fluxes
\balign{
\mflux^{\kk+\frac{1}{2}}=\mflux^{\kk-\frac{1}{2}}+\delmvapk-\left(  \sigma^{\kk-\frac{1}{2}} - \sigma^{\kk+\frac{1}{2}} \right)\sum_{\kk=0}^{\mathrm{N}} \delmvapk
}
Because $W^{N+\frac{1}{2}}\equiv0$, we can compute the other fluxes, and especially $W^{\frac{1}{2}}=\delmvap^0$, since $\sigma^{\frac{1}{2}}\equiv1$. Once the total mass fluxes are known, one can get the variation of the mixing ratios of the various tracers (generically called $q$ here) by considering the tracer mass budget
\balign{\label{dmq}
\delta \left( \massgk q^{\kk}\right)= q^{\kk-\frac{1}{2}} \mflux^{\kk-\frac{1}{2}}-q^{\kk+\frac{1}{2}} \mflux^{\kk+\frac{1}{2}} +\epsilon\, \delmvapk,
}
where $\epsilon$ is equal to 1 if the tracer considered is the vapor phase of the condensing gas, -1 if we consider the condensed phase, and 0 for every other tracers. The $q^{\kk-\frac{1}{2}}$ are the tracer mixing ratios transported through the interface $\sigma^{\kk-\frac{1}{2}}$. They are calculated using a "Van-Leer I" finite volume transport scheme\itref{Van77}.

Alternatively, the tracer mass conservation can be written exactly 
\balign{\label{dmq2}
\delta \left( \massgk q^{\kk}\right)= \left( \massgk + \delmassgk \right)\delta  q^{\kk} + q^{\kk}  \delmassgk ,
}
where $\delta  q^{\kk} $ is the correction to be applied to the tracer mixing ratios. Combining Eqs.\,(\ref{delmassW}), (\ref{dmq}), and (\ref{dmq2}), one can show that
\balign{\delta  q^{\kk} =&\ \  \frac{1}{\massgk + \delmassgk}\  \times
&\left[ \left(q^{\kk-\frac{1}{2}}- q^{\kk} \right) \mflux^{\kk-\frac{1}{2}}- \left(q^{\kk+\frac{1}{2}}- q^{\kk} \right) \mflux^{\kk+\frac{1}{2}} +\left(\epsilon- q^{\kk} \right)\, \delmvapk  \right],}
where the first terms represent transport through $\sigma$ levels and the last term corresponds to enrichment/depletion due to the variation of the gas mass. 

Finally, apart from tracers, the mass of gas that is advected through levels also transports some energy and momentum. Once the mass fluxes have been computed, these exchanges are easily computed using 
\balign{\delta  u^{\kk} =&\ \  \frac{1}{\massgk + \delmassgk}\ \left[ \left(u^{\kk-\frac{1}{2}}- u^{\kk} \right) \mflux^{\kk-\frac{1}{2}}- \left(u^{\kk+\frac{1}{2}}- u^{\kk} \right) \mflux^{\kk+\frac{1}{2}}  \right],}
where we have shown an example for the zonal speed $u$, but equations are the same for other quantities.

\section{Model validation at high temperatures}

To validate both the radiative transfer and the implementation of the moist adiabat in very hot atmospheres, we have developed a 1D "inverse climate modeling" version of our code, that has already been used in ref \ref{WP13}. In its spirit, this model is very similar to the codes developed in refs \ref{KPA84} and \ref{KRK13}.

For a given surface temperature and background gas surface pressure, the vertical thermal and water vapor profiles are integrated upward following a moist adiabatic lapse rate (\eq{lapse_rate}) until a fixed stratospheric temperature is reached (here 200K). Once the profile is built, the two stream radiative transfer routine used in our full 3D GCM is used to compute the outgoing thermal radiation and the bond albedo of the planet. In these calculations, the most important assumptions are that 1) the planet is spherically symmetric with a fixed surface albedo, 2) the atmosphere is always saturated in water vapor, and 3) the atmosphere is cloud free.

The results of this model are shown in Extended Data Fig.\,1 for a surface albedo of 0.25 to be comparable with the recent results of ref \ref{GRZ13}. We can see that the asymptotic behavior of the thermal flux as well as its quantitative asymptotic value (282\,W/m$^2$) are in good agreement with recently published similar models with up-to-date spectroscopic data\itref{KRK13}$^,$\itref{GRZ13}$^,$\itref{Pie10}. Our albedo calculations show a little discrepancy of less than 0.02 with respect to the relevant cases of ref \ref{GRZ13} (i.e. the pure water case and the transitional atmosphere case with 1b N$_2$ and preindustrial CO$_2$). These small differences are of little importance given the uncertainties on the primary mechanism controlling the albedo, i.e. the clouds.

\section{Intrinsic three-dimensional effects: the role of unsaturated regions}

To have more insight into the intrinsic differences between 1D and 3D simulations, we have performed a set of idealized simulations. In this numerical experiment, 
 we run our 3D model in a configuration without topography and with a uniform surface albedo (The effect of ice albedo is not taken into account). Furthermore, we do not take into account the radiative effect of clouds. Because we want to understand the role of the dynamically driven distribution of humidity, we treat the whole surface as an infinite reservoir of water. In these simulations, we are left with only one major free parameter (the background atmosphere being kept similar to the Earth present atmosphere, see above) which is the surface albedo that has been fixed to 0.22 to recover a mean surface temperature similar to the Earth one under the present insolation. 
 
The value of the mean surface temperature as a function of insolation is presented in Extended Data Fig.\,3a. For comparison, we have computed the surface temperature given by our 1D model for the same surface albedo. Because we do not take into account both cloud forcing and ice albedo in this set of 3D simulations, the only differences between the two models are due to dynamical effects and horizontal inhomogeneities in vapor and temperature distributions. One can readily see from Extended Data Fig.\,3a that the 3D model always predict lower temperatures than the 1D one. As a result, runaway greenhouse occurs at much higher insolation.

The explanation to this fact can be found in Extended Data Figs.\,3 and 4. Indeed, as observed on Earth, subtropical regions of the troposphere are unsaturated\itref{SIT10}$^,$\itref{Pie95}. As a result, these regions can emit more thermal flux than the asymptotic limiting flux (see Extended Data Fig.\,4b). At a given mean surface temperature, the global thermal flux can thus be larger than in the fully saturated 1D case, as can be seen in Extended Data Fig.\,3b. To retrieve comparable results with the 1D model, we have to decrease the relative humidity in the radiative transfer to 0.6 for present earth and $\sim$0.45 near the runaway greenhouse limit. Thus both the fact that subtropics are unsaturated and that they seem to get dryer under an increased insolation strongly stabilize the climate. Thus, as on Earth today, subtropics will continue to play the role of radiative fins\itref{Pie95}, deferring catastrophic consequences of a runaway greenhouse further in the future.

In the context of the runaway greenhouse, the effect of these unsaturations have been recognized in ref \ref{ITN02} where the authors performed a highly-idealized numerical experiment. They found that their 3D results could be mimicked in 1D by forcing a relative humidity of 0.6 or a little lower throughout the atmosphere. However, because of their use of a cloud-free, gray radiative transfer and of a very crude description of the water cycle, they were unable to make a quantitative assessment of the runaway greenhouse insolation limit for a more realistic cloudy, non-gray atmosphere. Indeed, this limit is directly determined by the gray opacity used, which is a free parameter. 
In terms of relative humidity, however, these authors find values and trends that are roughly similar to ours. Considering the very different setup involved (no ground thermal inertia, no seasonal or diurnal variations, ...), this suggests that, at least when looking at rough mean values, the mechanisms controlling the evolution of the relative humidity should be controlled by relatively simple processes\itref{SIT10}.
On the contrary, even if their very different temperature profiles, due to the gray approximation (especially their warm stratospheres), should not affect the runaway greenhouse threshold, they certainly prevent any robust conclusion considering water escape.

Finally, another result of our simulations is to confirm that both cloud and ice albedo feedbacks are destabilizing in the runaway greenhouse context. Indeed, because we keep the surface albedo fixed to a higher value than Earth surface today to mimic both the present effect of ice and clouds, this set of simulations can be seen as a case where clouds and ice are present but have no feedback. In this case, as visible on Extended Data Fig.\,3a, runaway occurs at a flux greater than $\sim\,400\,$W/m$^2$ which is much higher than the $\sim\,375\,$W/m$^2$ found when the cloud feedback is accounted for.

\section{Stratospheric temperatures in non-gray atmospheres}

One might be surprised by the fact that the stratospheric temperature found in our baseline model (see Fig.\,3) can be much colder than the skin temperature, $T_\mathrm{skin}=T_\mathrm{eq}/2^{1/4}$, where $T_\mathrm{eq}\equiv [(1-A) \Fs/ \sigma]^{1/4}$ is the equilibrium temperature, $A$ the bond albedo, $\Fs$ the mean insolation and $\sigma$ the Stefan-Boltzman constant.

The reason for the existence of a skin temperature in gray atmospheres is that there is, by construction, no radiative window by which the lower atmosphere and surface can cool to space without interacting with the upper atmosphere. The radiative equilibrium thus implies that the upward and downward emission from the optically thin upper atmosphere must both be equal to half the absorption of upward infrared flux, explaining the usual $2^{-1/4}$ factor.

As extensively discussed in ref \ref{Pie10}, in the real case, the radiation illuminating the upper atmosphere is depleted in the portion of the spectrum which is efficiently absorbed by the gas in the troposphere. The stratosphere, which, by definition, can efficiently absorb the upwelling infrared radiation only in this part of the spectrum, is thus poorly heated from below. However, it is still able to radiate efficiently, and is thus forced to cool to maintain radiative equilibrium balance. This does not threaten the global radiative balance of the whole atmosphere because even if the upwelling flux is small in the opaque regions of the troposphere, the surface and lower troposphere can cool through the transparent radiative windows.

A first quantitative, analytical estimate of the magnitude of this effect can be done using an idealized 1 band gas. As described in ref \ref{Pie10}, this simple model already shows that the ratio $T_\mathrm{skin}/T_\mathrm{eq}$ can be lower than inferred in the gray case. A much more comprehensive quantification of this phenomenon using an idealized (in the sense that they can arbitrarily choose the line shape) semi-analytical non-gray radiative model has been provided in ref \ref{AG72}. The main conclusions of their study was that in a non-gray atmosphere, the temperature increases with increasing mean opacity below a certain height and decreases with increasing mean opacity above that height. They also found that there is no lower limit to the temperature at the top of the atmosphere: it can arbitrarily  approach zero as the width of the lines is decreased.

While the aforementioned studies demonstrate that there is no theoretical paradox in having (arbitrarily) cold stratospheres, let us now turn to the validation of our more realistic radiative transfer model in the case of our ozone-free Earth-like planet. Recently, several different climate models (mainly addressing early Earth climate), have computed temperature profiles for an ozone-free atmosphere under present insolation. Among these, ref \ref{LTF13} find a minimum stratospheric temperature of 140-145K, and ref \ref{WT13} find it to be around 150\,K. In a more idealized, 1D setup, ref \ref{WP13} have published consistently computed 1D temperature profiles for N$_2$-CO$_2$-H$_2$O atmospheres and they also find stratospheric temperatures around 140-150\,K.
	Considering the fact that our estimation for the ozone-free stratospheric temperature ($\sim$140\,K) agrees well within the uncertainties with these numerous published results using different numerical models, we are confident that our non-gray radiative transfer model is suitable for the present application.

\clearpage
\begin{enumerate}
\setcounter{enumi}{28}
\item \label{HMB06} Hourdin, F. \textit{et al.} The LMDZ4 general circulation model: climate performance and sensitivity to parametrized physics with emphasis on tropical convection. \textit{Climate Dynamics}, \textbf{27}, 787-813 (2006).
\item \label{MY82} Mellor, G. L. \& Yamada, T. Development of a turbulence closure model for geophysical fluid problems. \textit{Reviews of Geophysics and Space Physics}, \textbf{20}, 851-875 (1982).
\item \label{RGB09} Rothman, L. S. \textit{et al.} The HITRAN 2008 molecular spectroscopic database. \textit{J. Quant. Spec. Radiat. Transf.}, \textbf{110}, 533-572 (2009).
\item \label{CKD89} Clough, S., Kneizys, F. \& Davies, R. Line shape and the water vapor continuum. \textit{Atmospheric Research}, \textbf{23}, 229-241 (1989).
\item \label{BF86} Borysow, A. \& Frommhold, L. Collision-induced rototranslational absorption spectra of N2-N2 pairs for temperatures from 50 to 300 K. The \textit{Astrophys. J.}, \textbf{311}, 1043-1057 (1986).
\item \label{RGR11} Richard, C. \textit{et al.} New section of the hitran database: Collision-induced absorption (cia). \textit{J. Quant. Spec. Radiat. Transf.} (2011).
\item \label{TMA89} Toon, O. B., McKay, C. P., Ackerman, T. P. \& Santhanam, K. Rapid calculation of radia- tive heating rates and photodissociation rates in inhomogeneous multiple scattering atmospheres. \textit{J. Geophys. Res.}, \textbf{94}, 16287-16301 (1989).
\item \label{HT74} Hansen, J. E. \& Travis, L. D. Light scattering in planetary atmospheres. \textit{Space Sci. Rev.}, \textbf{16}, 527-610 (1974).
\item \label{Buc95} Bucholtz, A. Rayleigh-scattering calculations for the terrestrial atmosphere. \textit{Appl. Opt.}, \textbf{34}, 2765 (1995).
\item \label{LL91} Le Treut, H. \& Li, Z.-X. Sensitivity of an atmospheric general circulation model to prescribed sst changes: feedback effects associated with the simulation of cloud optical properties. \textit{Climate Dynamics}, \textbf{5}, 175-187 (1991).
\item \label{MW67} Manabe, S. \& Wetherald, R. T. Thermal Equilibrium of the Atmosphere with a Given Distribution of Relative Humidity. \textit{Journal of Atmospheric Sciences}, \textbf{24}, 241-259 (1967).
\item \label{Fri07} Frierson, D. M. W. The Dynamics of Idealized Convection Schemes and Their Effect on the Zonally Averaged Tropical Circulation. \textit{Journal of Atmospheric Sciences}, \textbf{64}, 1959 (2007).
\item \label{BLB95} Boucher, O., Le Treut, H. \& Baker, M. B. Precipitation and radiation modeling in a general circulation model: Introduction of cloud microphysical processes. \textit{J. Geophys. Res.}, \textbf{100}, 16395-16414 (1995).
\item \label{FML06} Forget, F., Hourdin, F. \& Talagrand, O. CO2 Snowfall on Mars: Simulation with a General Circulation Model. \textit{Icarus}, \textbf{131}, 302-316 (1998).
\item \label{Van77}  Van Leer, B. Towards the ultimate conservative difference scheme: IV. A new approach to numerical convection. \textit{J. Comput. Phys.}, \textbf{23}, 276-299 (1977).
\item \label{LTF13} Le Hir, G., Teitler, Y., Fluteau, F., Donnadieu, Y. \& Philippot, P. The faint young Sun problem revisited with a 3-D climate-carbon model - Part 1. \textit{Clim. Past Discuss.}, 9, 1509-1534, (2013).
\item \label{WT13} Wolf, E. \& Toon, O., B. Hospitable Archean Climates in a General Circulation Model. \textit{Astrobiology}. \textbf{13}, 656-673, (2013).
\end{enumerate}

\clearpage

%\cite{HMB06,MY82,RGB09,CKD89,BF86,RGR11,TMA89,HT74,Buc95,LL91,Fri07,BLB95,FML06,FHT98,MW67}

%\bibliography{../biblio} 
%\bibliographystyle{astron} 
%\bibliographystyle{naturemag} 
%\bibliographystyle{aa} 

\clearpage

%\clearpage

\section*{Extended Data Figure}

\begin{figure}[htb] %  figure placement: here, top, bottom, or page
 \centering
  \subfigure{ \includegraphics[scale=.6,trim = 0.cm 0.cm .0cm 0cm, clip]{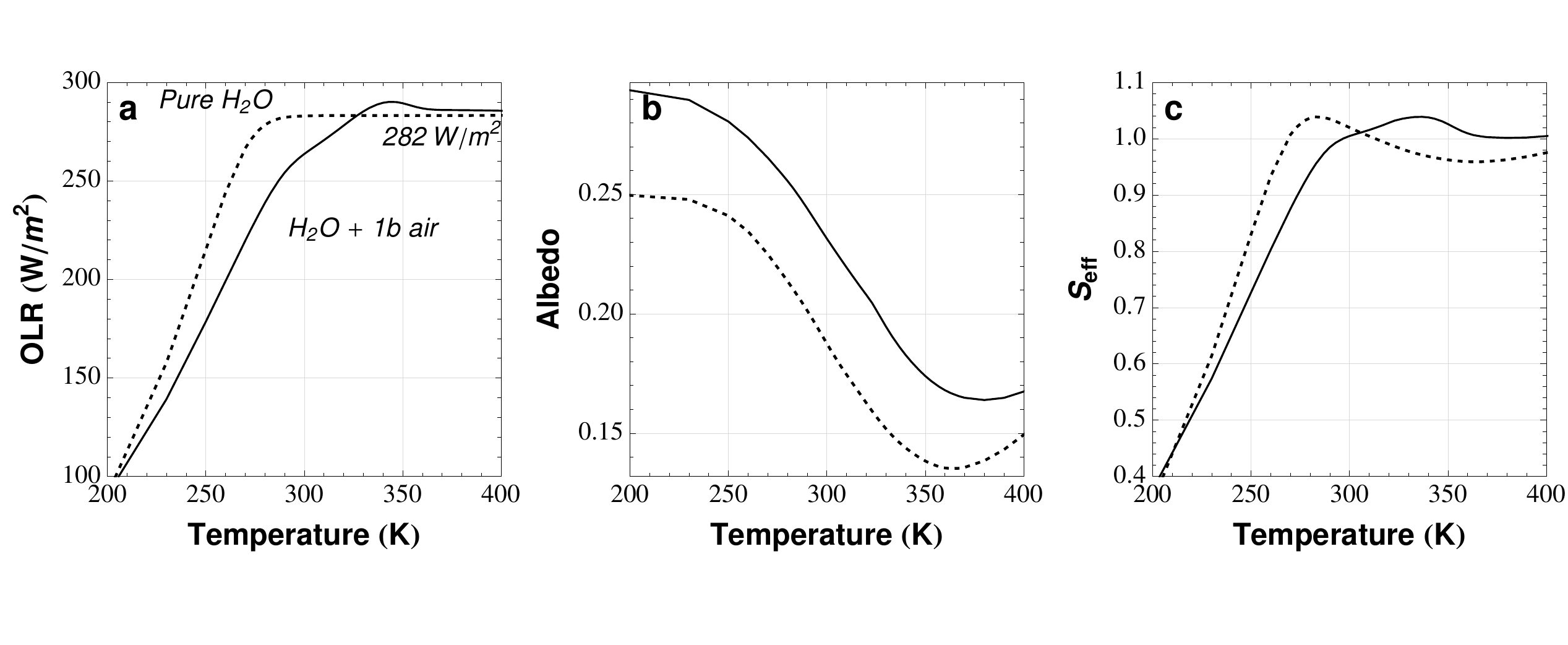} }
\vspace{-1cm}
 \caption*{
\textbf{Extended Data Figure 1: Validation of the radiative transfer model at high temperature}. Emitted thermal flux (a), bond albedo (b), and effective solar constant (with respect to the current solar constant; c) dependence on surface temperature with a 1D version of our GCM in the "reverse climate modeling" mode\itref{KPA84}$^,$\itref{KRK13}. The dashed curve is the pure water case, and the solid curve is the case with a 1 bar N$_2$ background atmosphere including 376\,ppmv of CO$_2$. The surface albedo is 0.25.
}
 \label{fig:kcm1d}
\end{figure}
\clearpage

\vspace{-5cm}

\begin{figure*}[htb] %  figure placement: here, top, bottom, or page
 \centering
 \subfigure{ \includegraphics[scale=1.,trim = 0.cm 0.cm .0cm 0cm, clip]{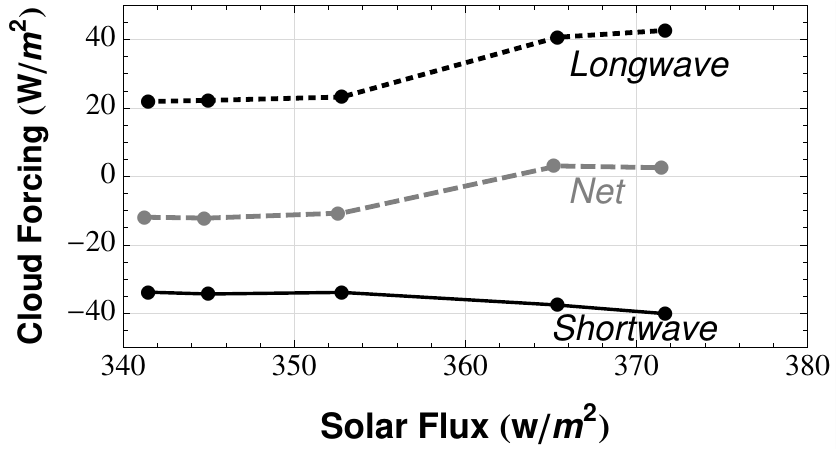} }
 \caption*{
\textbf{Extended Data Figure 2:
Evolution of the cloud radiative forcing with the mean solar incoming flux for the scenario with fixed cloud particle radii}. Solid, dotted and dashed curves are the shortwave, longwave and net radiative cloud forcing. Although less simulations have been run, the changes in the value of the slopes around 353 and 365\,W/m$^2$ seem to have the same origin as the behavior change seen in Fig.\,2c (although they occur at different fluxes). These changes in cloud behavior might be due to the disappearance of both permanent ice caps (at lower fluxes) and seasonal snow covers (at higher fluxes).
}
 \label{fig:CRFradfix}
\end{figure*}

\begin{figure}[htbp] %  figure placement: here, top, bottom, or page
 \centering
 \subfigure{ \includegraphics[scale=.8,trim = 0.cm 0.cm .0cm 0cm, clip]{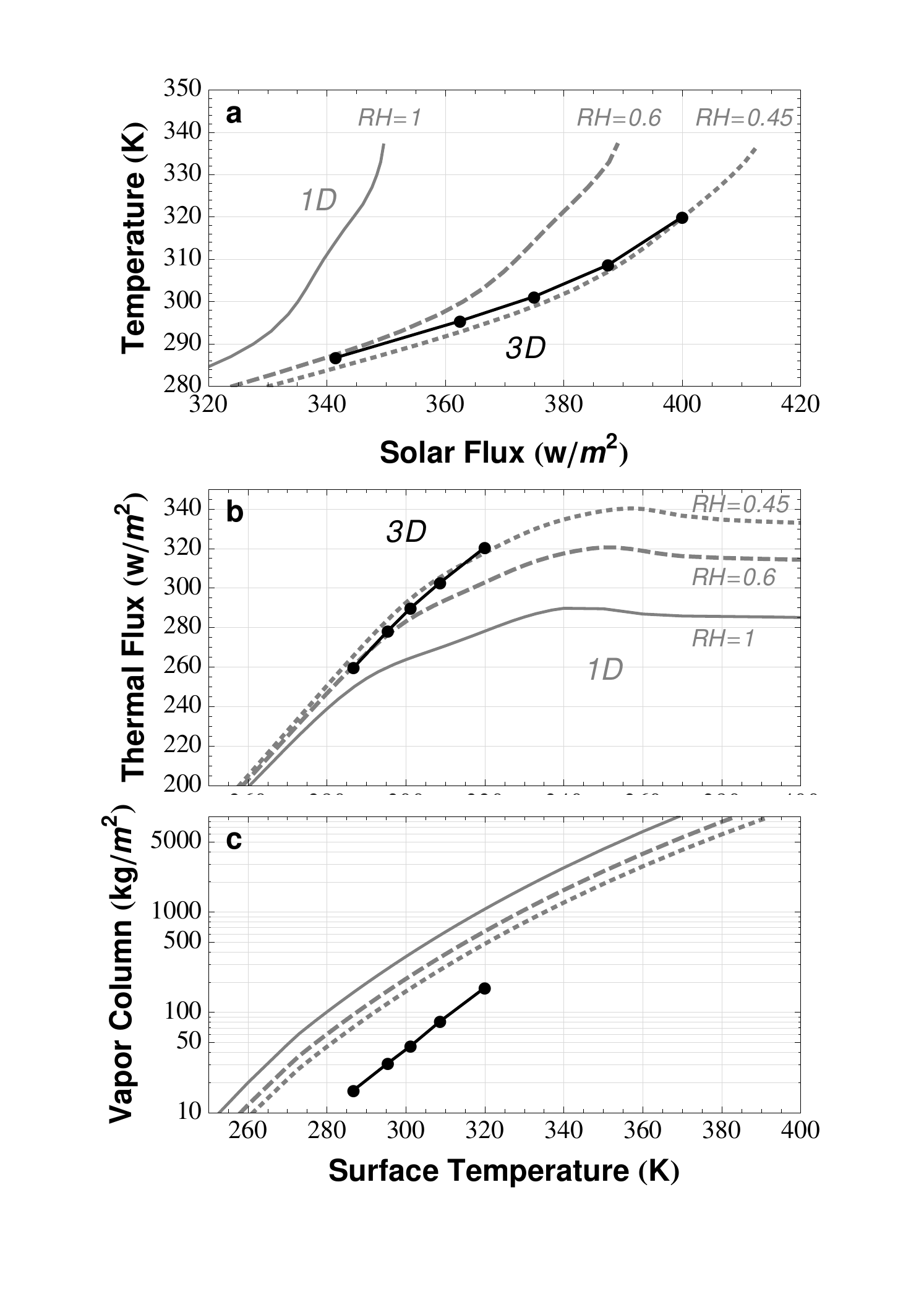} }
 \caption*{
\textbf{Extended Data Figure 3:
Comparison between 1D and 3D cloud-free aquaplanet simulations}. \textbf{a}, Mean surface temperature as a function of incoming stellar flux. \textbf{b}, Emitted thermal flux dependence on surface temperature. \textbf{c}, Water vapor column as a function os surface temperature. In all panels, filled dots stand for the idealized 3D set of aquaplanet simulations, and gray curves stand for the 1D model. In both cases, a uniform surface albedo of 0.22 is used. In the 1D case, the relative humidity in the radiative transfer is forced to be 1, 0.6 and 0.45 (solid, dashed and dotted curves respectively).
}
 \label{fig:idealized1}
\end{figure}

\begin{figure}[htbp] %  figure placement: here, top, bottom, or page
 \centering
  \subfigure{ \includegraphics[scale=.8,trim = 0.cm 0.cm .0cm 0cm, clip]{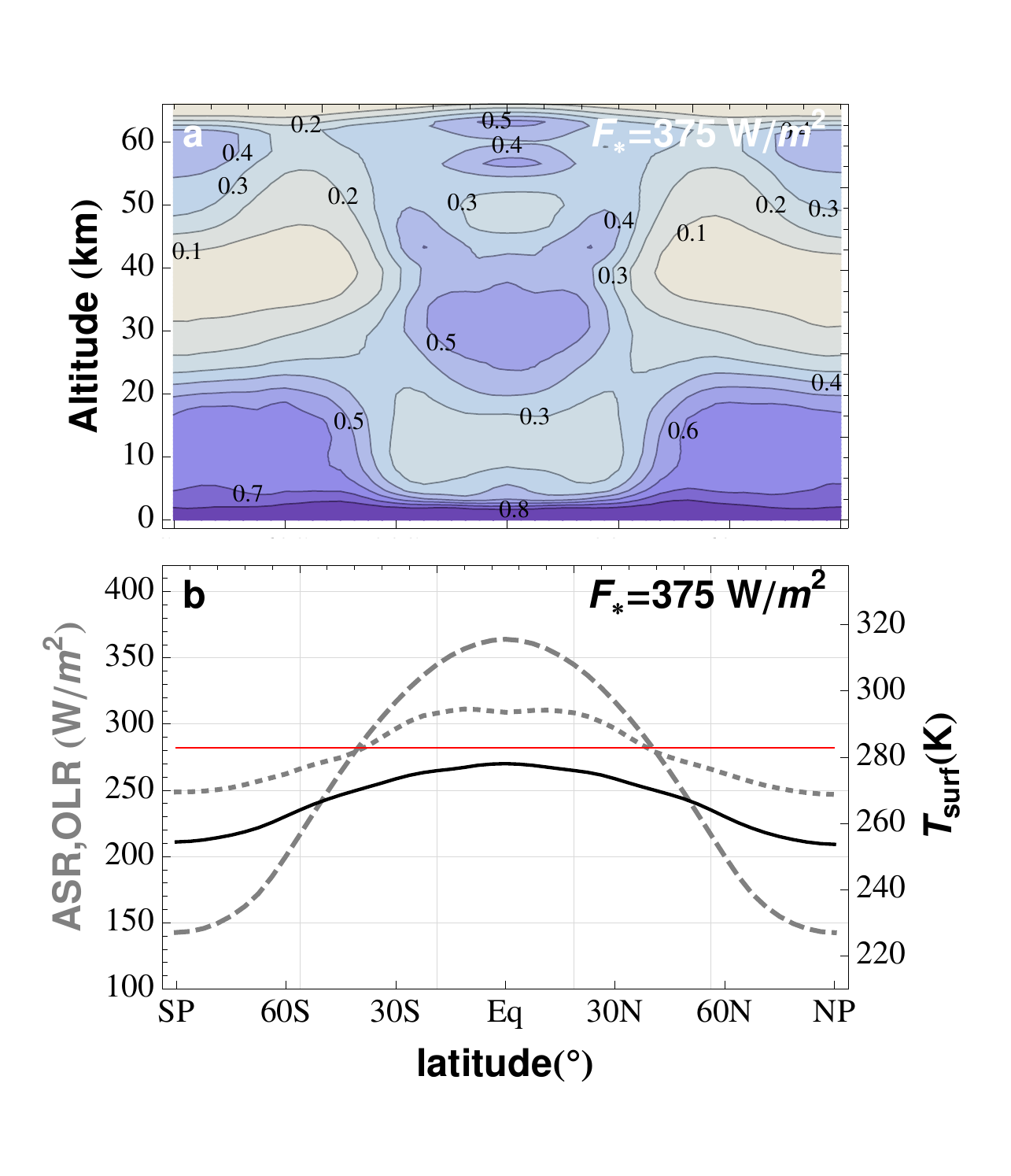} }
 \caption*{
\textbf{Extended Data Figure 4: 
Relative humidity and radiative budget for an idealized, cloud-free aquaplanet.}
\textbf{a}, Annually and zonally averaged relative humidity in a latitude altitude plane. \textbf{b}, Absorbed (gray dashed) and emitted (gray dotted) flux and surface temperature (solid black) distribution with latitude (annually and zonally averaged). The red line is the asymptotic limit infrared flux for a saturated atmosphere. Results are shown for the case receiving 375\,W/m$^2$.
}
 \label{fig:idealized3}
\end{figure}

\begin{figure}[htbp] %  figure placement: here, top, bottom, or page
 \centering
  \subfigure{ \includegraphics[scale=.4,trim = 0.cm 0.cm .0cm 0cm, clip]{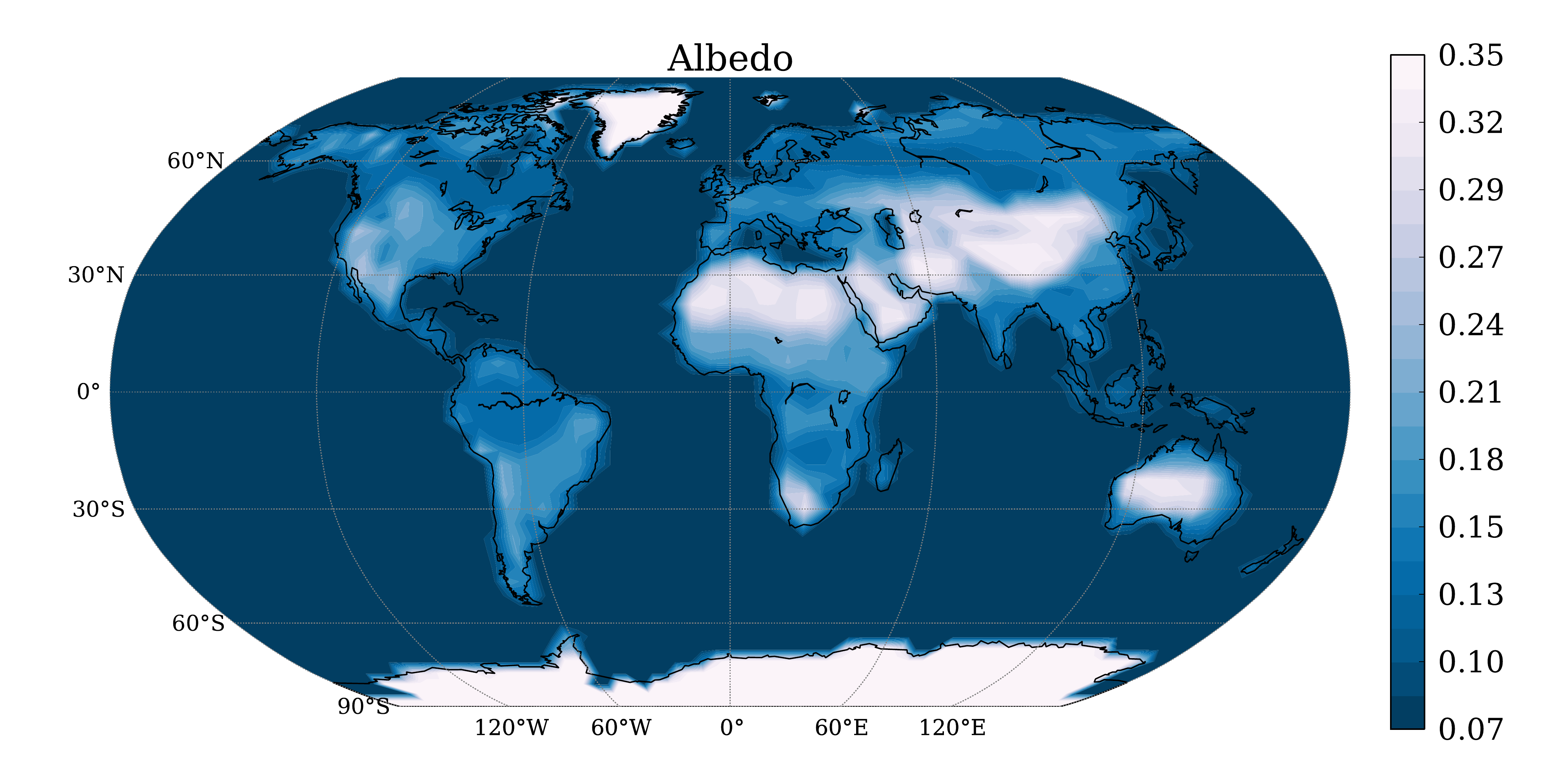} }
 \caption*{
\textbf{Extended Data Figure 5:
Surface albedo map used for the Earth baseline case}. This map does not include the effect of the ice albedo which is computed directly by the GCM. The albedo of Greenland and Antartica, in particular, was set to 0.35. The altitude of these regions was, however, left unchanged, explaining in part the temperature contrasts around these areas in Fig.\,1. 
}
 \label{fig:albedomap}
\end{figure}

\clearpage

\end{document}